\documentclass{revtex4}
\usepackage {amssymb}
\usepackage {amsmath}
\usepackage {epsfig}
\usepackage{graphicx}
\usepackage {hyperref}

\def\be {\begin{equation}}
\def\ee {\end{equation}}
\def\beq{\begin{eqnarray}}
\def\eeq{\end{eqnarray}}

\begin{document}

\markboth{B.C. L\"{u}tf\"{u}o\={g}lu and F. Ta\c{s}k\i{i}n}
{Renormalization Group Analysis of a G\" ursey Model Inspired
Field Theory II}

\title{Renormalization Group Analysis of a G\" ursey Model Inspired
Field Theory II}%

\author{B.C. L\"{u}tf\"{u}o\={g}lu$^1$}%
\email{bcan@itu.edu.tr}%

\author{F. Ta\c{s}k\i n$^{1,2}$ }%
\email{taskinf@itu.edu.tr}%
\affiliation{$^1$ Department of Physics, Istanbul Technical
University, Istanbul, Turkey \\$^2$ Department of Physics, Erciyes
University, Kayseri, Turkey.}

\date{\today}%

\begin{abstract}
Recently  a model, which is equivalent to the scalar form of G\"
ursey model, is shown to be a nontrivial field theoretical model
when it is gauged with a $SU(N)$ field. In this paper we study
another model that is equivalent to the vector form of the
G\"{u}rsey model. We get a trivial theory when it is coupled with
a scalar field. This result changes drastically when it is coupled
with an additional $SU(N)$ field. We find a nontrivial field
theoretical model under certain conditions.
\end{abstract}%

\maketitle%

\section{Introduction}

Historically, there has always been a continuing interest in
building  nontrivial field theoretical models. A while ago it was
shown that perturbative expansions are not adequate in deciding
whether a model is nontrivial or not. Baker et al. showed that the
$\phi^{4}$ theory, although perturbatively nontrivial, went to a
free theory as the cutoff was lifted  in four dimensions
\cite{ba_ki_79,ba_ki_81}. Continuing research is going on this
subject \cite{kl_06}. Alternative methods become popular.
Renormalization group (RG) methods are the most commonly used one.
They were first introduced by Wilson et al. \cite{wi_ko_74}.
Another method is using exact RG algorithm which were proposed by
Polchinski \cite{po_84}. Recent studies gave important insights on
both methods \cite{so_07_40_5733,so_07_40_9675,ig_it_so_07}.

Another endeavor is building a model of nature using only
fermions. Here all the observed bosons are constructed as
composites of these ingredient spinors. In solid state physics,
electrons come together to form bosonic particles \cite{mi_93,
ba_co_sc_57}. Historically, the first work on models with only
spinors goes back to the work of Heisenberg \cite{he_54}. Two
years later G\"{u}rsey proposed his model as a substitute for the
Heisenberg model \cite{gu_56}. This G\"{u}rsey's spinor model is
important since it is conformally invariant classically and has
classical solutions \cite{ko_56} which may be interpreted as
instantons and merons \cite{ak_82}, similar to the solutions of
pure Yang-Mills theories in four dimensions \cite{fu_sh_90}. This
original model can be generalized to include vector, pseudovector
and pseudoscalar interactions.

We have worked on different forms of the G\" ursey model
\cite{ho_lu_06,ho_ta_07,ho_lu_ta_07} using the earlier works
\cite{ak_ar_du_ho_ka_pa_82-34,ak_ar_du_ho_ka_pa_82-41,ak_ar_ho_pa_83,ar_ho_83,ar_ho_ka_85}
as a starting point. In those references it was claimed that a
polynomial lagrangian could be written equivalently to G\" ursey's
non-polynomial lagrangian. Recently it is shown that they are
equivalent only in a naive sense \cite{ho_lu_06,ho_ta_07}. In
\cite{ho_lu_06}, using perturbative methods, we showed that only
composite particles took part in physical processes whereas the
constituent fields  did not interact with each other. Recently in
\cite {ho_lu_ta_07}, we showed that, when this model is coupled to
a constituent $U(1)$ gauge field, we were mimicking a gauge
Higgs-Yukawa (gHY) system, which had the known problems of the
Landau pole, with all of its connotations of triviality. There,
our motivation was the famous Nambu-Jona-Lasinio model
\cite{na_jo_61}, which was written only in terms of spinor fields.
This model was shown to be trivial \cite{ko_ko_94,zi_89}. Recent
attempts to gauge this model to obtain a nontrivial theory are
given in references
\cite{ha_ki_ku_na_94,ao_mo_su_te_to_99,ao_mo_su_te_to_00,ku_te_99,ko_ta_ya_93}.

The essential point in our analysis is the factor of $\epsilon$ in
the composite propagator \cite{ho_lu_06,ho_ta_07}. This main
difference makes many of the diagrams convergent when the cutoff
is removed. Consequently, we find that we can construct a
nontrivial model from the scalar G\" ursey model when a
non-Abelian gauge field is coupled to the fermions
\cite{ho_lu_07}. In this paper we will investigate the vector form
of the G\" ursey model. Here we will closely follow the line of
discussion followed in the references
\cite{ha_ki_ku_na_94,ho_lu_07}.

This article is organized as follows. In the next section we
describe the vector form of the G\" ursey like model. There we
derive the composite vector field propagator. In section 3, we
couple a constituent scalar field to our model and discuss the new
results. Then we solve the renormalization group equations (RGE's)
and find a Landau pole in the solution. In section 4, we introduce
another field, a non-Abelian gauge field to the model. In the
subsections we write the new RGE's and derive the solutions by
using some RG invariants. We discuss some limiting cases of the
coupling constant solutions before giving the criteria's of the
nontriviality condition in section 5. Then we find the fixed point
solutions. In the following subsections we analyze the solutions
of the coupled equations and find their asymptotic behaviors. The
final section is devoted to conclusions.

\section{The Model}

The vector form of the pure spinor G\" ursey model
\cite{ak_ar_ho_pa_83} is given as

\beq %
    L=\overline{\psi}\left(i\partial\!\!\!/ -ig\partial
   \!\!\!/g^{-1}-m\right)\psi+\alpha \left[(\overline{\psi}
   \gamma_{\mu}\psi)(\overline{\psi}\gamma^{\mu}\psi)\right]^{2/3}.
    \label{gursey lagrangian}
\eeq %

Here only the spinor fields have kinetic part. The $g$ field is a
pure gauge term to restore the local gauge symmetry, when the
spinor field is transformed. This non-polynomial Lagrangian has
been converted to an equivalent polynomial form by introducing
auxiliary fields $\lambda_{\mu}$ and $G_{\mu}$ in
\cite{ak_ar_ho_pa_83}. The constrained Lagrangian in the
polynomial form is given as

\beq %
 L_{c}=\overline{\psi}\left[i\partial \!\!\!/
   -ig\partial \!\!\!/g^{-1}+e(G\!\!\!/
   +\lambda\!\!\!/)-m\right]\psi
   -e^4\lambda_{\mu}G^{\mu}G^{2}
   +\mbox{ghost terms}. \label{s yenilagran} %
\eeq %

Recently it was shown that this equivalence should be taken only
"naively" \cite{ho_ta_07}. This expression contains two constraint
equations, obtained from writing the Euler-Lagrange equations for
the auxiliary fields. Hence it should be quantized by using the
constraint analysis $\grave{a}$ \textit{la} Dirac \cite{di_64}.
This calculation is performed  using the path integral method. We
find out that one can write the effective Lagrangian as
\beq %
   L_{eff}=\overline{\psi}\left[i\partial \!\!\!/ -ig\partial
  \!\!\!/ g^{-1}+e\left( G\!\!\!/+\lambda \!\!\!/
  \right)-m\right]\psi -e^4\lambda_{\mu}G^{\mu}G^{2}+
  \overline{w}^{\mu}(g_{\mu\nu}G^2 + 2G_{\mu}G_{\nu})w^{\nu}.
  \label{etkinlagran} %
\eeq
Here $\overline{w}^\mu$ and $w^{\nu}$ are the ghost fields. With a
suitable redefinition of the fields the effective action  can be
given as
\beq %
    S_{eff}=Tr\ln\left(i\partial \!\!\!/+eJ\!\!\!/+m\right)
    + \int dx^4 \left[e^4\left(J_{\mu}J^{\mu}J_{\lambda}J^{\lambda}\right)
    +\mbox{other terms} \right], \label{eff_action}%
\eeq %
where $ J_{\mu}=-ig\partial_{\mu} g^{-1}+ G_{\mu}+ \lambda_{\mu}$.
The second derivative of the effective action with respect to the
$J_{\mu}$ field gives us the induced inverse propagator as
\beq %
  \left . \frac{\partial^2 S_{eff}}{\partial J_{\mu}\partial
   J_{\nu}}\right |_{J_{\mu}=0}= -\frac{g^2}{3\pi^2}\left(q_{\mu}q_{\nu}-g_{\mu\nu}q^2\right)
   \left[\frac{1}{\epsilon}+\mbox{finite part}\right]. %
\eeq %
Here dimensional regularization is used for the momentum integral
and $\epsilon = 4-n$. All the other fields not shown in this
expression, including ghost fields arising from the constrained
equations, decouple from the model. The only remaining fields are
the spinors and the $J_{\mu}$ field. This procedure is explicitly
carried out in \cite{ho_ta_07}. In the Feynman gauge the
propagator of the composite vector field can be written as
$\epsilon\frac{g^{\mu\nu}}{p^2}$ where the spinor propagator is
the usual Dirac propagator in the lowest order.

Although the original Lagrangian does not have a kinetic term for
the vector field, one loop corrections generate this term and make
this composite field as a dynamical entity like it is done in
\cite{ho_lu_06}, where the composite vector field is replaced by
composite scalar field. In the literature there are also other
similar models with differential operators in the interaction
Lagrangian \cite{am_ba_da_ve_81}.

In reference \cite{ho_ta_07}, the contributions to the fermion
propagator at higher orders were investigated by studying the
Dyson-Schwinger equations for the two point function. We found
that there is a phase which has no additions to the existing
fermion  mass. %
%
%
\section{Coupling with A Scalar Field}
We may add a constituent complex scalar field to the model and
investigate the consequences of this addition. Our motivation is
the work of  Bardeen et al. \cite {ba_le_lo_86, le_lo_ba_86}. When
they added a vector field to the Nambu-Jona-Lasinio model, a
complementary procedure to our work, they got interesting results.
Since we already have a composite vector field, we can couple a
massless scalar field which has its kinetic term, a self
interacting term with coupling constant $a$ and an interaction
term  with new coupling constant $y$ in the Lagrangian. Then the
effective Lagrangian becomes
\beq %
   L_{eff}=\overline{\psi}\left[i\partial \!\!\!/ -ig\partial
  \!\!\!/ g^{-1}+e\left( G\!\!\!/+\lambda \!\!\!/
  \right)-m\right]\psi -e^4\lambda_{\mu}G^{\mu}G^{2}+
  \overline{w}^{\mu}(g_{\mu\nu}G^2 + 2G_{\mu}G_{\nu})w^{\nu}+
  \frac{\partial_{\mu}\phi\partial^{\mu}\phi}{2}-
  \frac{a}{4}\phi^{4}-y\overline{\psi}\phi\psi.
  \label{etkinlagran+skalar1} %
\eeq %
Since the $G_{\mu}$, $\lambda_{\mu}$ and ghost fields decouple,
this Lagrangian reduces to the effective expression given below.
\beq %
   L_{eff}=\overline{\psi}\left(i\partial \!\!\!/ +eJ\!\!\!/-
   y\phi-m\right)\psi -e^4J^{4}+ \frac{\partial_{\mu}\phi
   \partial^{\mu}\phi}{2}- \frac{a}{4}\phi^{4}.
  \label{etkinlagran+skalar2} %
\eeq %
If our fermion field had a color index $i$ where $i=1...N$, we
could perform an 1/N expansion to justify the use of only ladder
diagrams for higher orders for the scattering processes. Although
in our model the spinor has only one color, we still consider only
ladder diagrams anticipating that one can construct a variation of
the model with N colors. In the following subsection we summarize
the changes in our results for this new model.
\subsection{New Results and Higher Orders}
In the model described in reference \cite{ho_ta_07}, it is shown
that only composites can scatter from each other with a finite
expression, due to the presence of  $\epsilon$ in the composite
vector propagator.  There is also a tree-diagram process where the
spinor scatters from a composite particle, a Compton-like
scattering, with a finite cross-section. This diagram can be
written in the other channel, which can be interpreted as spinor
production out of vector particles. Note that in the original
model the four spinor kernel was of order $\epsilon $. The lowest
order diagram, vanishes due to the presence of the composite
vector propagator. In higher orders this expression can be written
in the quenched ladder approximation \cite{mi_93}, where the
kernel is separated into a vector propagator with two spinor legs
joining the proper kernel. If the proper kernel is of order
$\epsilon$, the loop involving two spinors and a vector propagator
can be at most finite that makes the whole diagram in first order
in $\epsilon$. This fact shows that there is no nontrivial
spinor-spinor scattering in the original model.

\begin{figure}[h]
\begin{center}
$\begin{array}{c@{\hspace{1cm}}c@{\hspace{5mm}}c}
 \multicolumn{1}{l}{\mbox{\bf }}&
 \multicolumn{1}{l}{\mbox{\bf }}&
 \multicolumn{1}{l}{\mbox{\bf }}\\
 [-0.53cm]
 \epsfxsize=20mm \epsffile{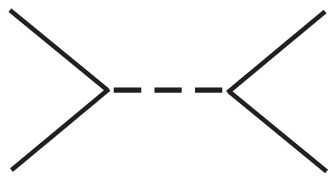}&
 \epsfxsize=18mm \epsffile{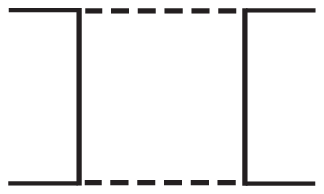}&
 \epsfxsize=22mm \epsffile{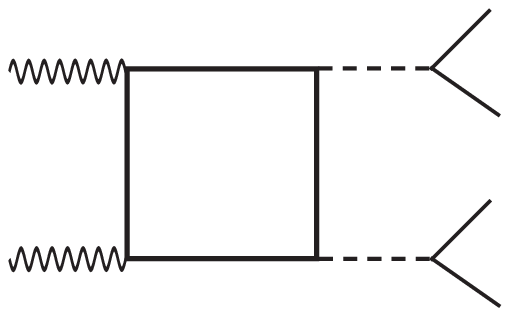} \\
 [0.4cm]
 \mbox{\bf (a)} &
 \mbox{\bf (b)} &
 \mbox{\bf (c)}
 \end{array}$
\end{center}
\caption{(a) Two spinor scattering through the scalar particle
channel, (b) Higher order diagram for two spinor scattering,
\hspace{8mm}(c) Spinor production from scattering of composite
vectors.} \label{fig789}
\end{figure}

These results changes drastically with scalar field coupling. Two
fermion scattering is now possible due to the presence of the
scalar field instead of vector field channel. In lowest order this
process goes through the tree diagram given in Figure
\ref{fig789}.a. At the next higher order the box diagram with two
spinors and two scalar particles, Figure \ref{fig789}.b,  is
finite from dimensional analysis. If the scalar particles are used
as intermediaries, the spinor production from scattering of
composite vector particles becomes possible as shown in  Figure
\ref{fig789}.c where the dotted, straight and wiggly lines
represent scalar, spinor and composite vector particles,
respectively.

\subsection{Renormalization Group Equations and Solutions}
In reference \cite{ho_ta_07}, it is widely discussed that the
$<{\overline{\psi}}\psi J_\mu>$ vertex and the spinor box diagram
give finite results. The higher diagrams do not change this
result, since each momentum integration is accompanied by an
$\epsilon$ term in the composite vector propagator. Therefore,
there is no need for infinite coupling constant renormalization.

\begin{figure}[h]
\begin{center}
$\begin{array}{c@{\hspace{1cm}}c@{\hspace{5mm}}c@{\hspace{5mm}}c}
 \multicolumn{1}{l}{\mbox{\bf }}&
 \multicolumn{1}{l}{\mbox{\bf }}&
 \multicolumn{1}{l}{\mbox{\bf }}&
 \multicolumn{1}{l}{\mbox{\bf }}\\
 [-0.53cm]
 \epsfxsize=13mm \epsffile{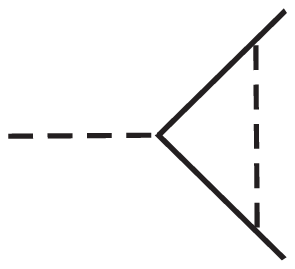} &
 \epsfxsize=13mm \epsffile{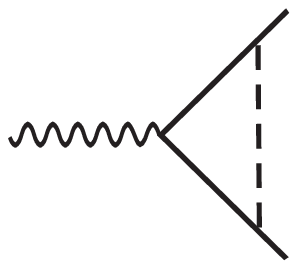} &
 \epsfxsize=20mm \epsfysize=8mm \epsffile{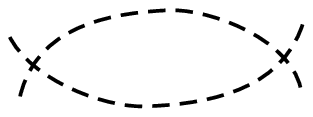} &
 \epsfxsize=16mm \epsffile{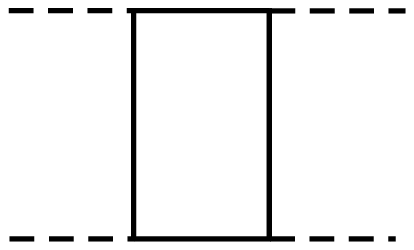} \\
 [3mm]
 \mbox{\bf (a)} &
 \mbox{\bf (b)} &
 \mbox{\bf (c)} &
 \mbox{\bf (d)}
 \end{array}$
\end{center}
\caption{The three coupling constant corrections in one loop. }
\label{fig456}
\end{figure}

In the new model where a massless scalar field is added, all the
three coupling constants are renormalized. One can write the first
order RGE's for these coupling constants, similar to the analysis
in \cite{ha_ki_ku_na_94}. We take $\mu_0$ as a reference scale at
low energies, $t=ln (\mu/\mu_0)$, where $\mu$ is the
renormalization point.
\beq
 16\pi^2\frac{d}{dt}y(t) &=& A y ^3(t),\label{y3} \\
 16\pi^2\frac{d}{dt}e(t) &=& B e(t) y^2(t),\label{eg2}\\
 16\pi^2\frac{d}{dt}a(t) &=& C a^2(t)-D y^4(t) \label{g4}.
\eeq
Here $A$, $B$, $C$ and $D$ are positive numerical constants. We
find out that Yukawa and $<{\overline{\psi}}\psi J_\mu>$ vertices
have only scalar correction. The composite vector correction to
these vertices are finite due to the $\epsilon$ in the propagator.
Therefore, our equations differ from those in reference
\cite{ho_lu_07,ha_ki_ku_na_94}. These processes are illustrated in
diagrams shown in Figure \ref{fig456}.

The RGE's have the immediate solutions
\beq
 y^2(t)&=&\frac{y_0^2}{Z(t)}, \\
 e(t)&=&e_0Z(t)^{-B/2A}, \\
 a(t)&=& \frac{A\pm\sqrt{A^2+CD}}{C}\frac{y_0^2}{Z(t)},
\eeq
where $Z(t)=1-\frac{Ay_0^2}{8\pi^2}t$.

The main problem of models with U(1) coupling, namely the Landau
pole, is expected to make our new model a trivial one. We expect
that coupling to a non-Abelian gauge theory will remedy this
defect by new contributions to the RGE's. Thus, obtaining a
nontrivial model will be possible. Coupling to a non-Abelian gauge
field will also give us more degrees of freedom in studying the
behavior of the beta function. This may allow us to find the
critical number of gauge and fermion fields to obtain a zero of
this function at nontrivial values of the coupling constants of
the model.
\section{Coupling with a Non-Abelian Field}
In this section we consider our model with $SU(N_{C})$ gauge field
interaction, where the spinors have $N_{f}$ different flavors.
Although we study in the leading order of $\frac{1}{N_{C}}$
expansion, where all the planar diagrams contribute to the RGE's,
we are interested in the high-energy asymptotic region where the
gauge coupling is perturbatively small;
$\frac{g^{2}N_{C}}{4\pi}\ll 1$. However, the number of fermions is
in the same order as $N_{C}$. Only $n_{f}$ fermions have a
degenerate large Yukawa coupling. We start with the effective
Lagrangian
\beq %
   L_{eff}=\sum_{i=1}^{N_{f}}\overline{\psi}_{i}\left(iD \!\!\!\!/ +eJ\!\!\!/
   -m\right)\psi_{i}-e^4J^{4}+ \frac{\partial_{\mu}\phi
   \partial^{\mu}\phi}{2}- \frac{a}{4}\phi^{4}-\sum_{i=1}^{n_{f}}
   \overline{\psi}_{i}y\phi\psi_{i}-\frac{1}{4} Tr [F_{\mu \nu} F^{\mu \nu}]
   + L_{\mbox{ghost}}+ L_{\mbox{g.f.}}.
  \label{etkinlagran+skalar+gauge} %
\eeq %
The gauge field belongs to the adjoint representation of the color
group $SU(N_C) $ where $D_{\mu}$ is the color covariant
derivative. $y$, $a$, $e$ and $g$ are the Yukawa, quartic scalar,
composite vector and gauge coupling constants, respectively.

There are two kind of ghost fields in the model. The first one,
which comes from the composite constraints, decouples from our
model \cite{ak_ar_ho_pa_83,ak_ar_du_ho_ka_pa_82-41}. The second
one, coming from the gauge condition on the vector field, do not
decouple and contribute to the RGE's in the usual way.
\subsection{Renormalization Group Equations and Solutions}
In this subsection we will analysis the RGE's in the leading order
of the approximation given above.  In the one loop approximation
the RGE's are
\beq
 16\pi^2\frac{d}{dt}g(t) &=& -A g^3(t),\label{5g3} \\
 16\pi^2\frac{d}{dt}y(t) &=& B y^3(t)-Cy(t)g^2(t),\label{5y3}\\
 16\pi^2\frac{d}{dt}e(t) &=& D e(t)y^2(t)-Ee(t)g^2(t),\label{5eg2}\\
 16\pi^2\frac{d}{dt}a(t) &=& F a(t)y^2(t)-Gy^4(t)
 \label{5L4}.
\eeq
Here $A$, $B$, $C$, $D$, $E$, $F$ and $G$ are positive constants.

In the RGE's we see that the diagrams, where the composite vector
field takes part, are down by order of $\epsilon$. Therefore we do
not have contributions proportional to $e^2(t)$, $e^3(t)$,
$y(t)e^2(t)$ and $g(t)e^2(t)$. Also we neglect the scalar loop
contribution to the gauge coupling $g(t)$, similar to the work of
\cite{ha_ki_ku_na_94}.

The solutions of the first RG equation (\ref{5g3}) can be given as %
\beq%
g^{2}(t)=g_{0}^{2}\Bigg(1+\frac{A\alpha_0}{2\pi}t\Bigg)^{-1},
\label{g_nin_cozumu}%
\eeq%
where $\alpha_0=\frac{g_{0}^2}{4\pi}$. We define  %
\beq
\eta(t)\equiv\frac{\alpha(t)}{\alpha_{0}}\equiv\frac{g^2(t)}{g_0^2},
\eeq %
where $g_0=g(t=0)$ which is the initial value at the reference
scale $\mu_0$. For the solution of the second RG equation
(\ref{5y3}), we can propose a RG invariant $H(t)$ as
\beq
H(t)=-\eta^{-1+C/A}(t)\left[1-\frac{C-A}{B}\frac{g^2(t)}{y^2(t)}\right].
\eeq %
Since $H(t)$ is a constant, we  call it $H_0$. Then, the solution
of the Yukawa coupling constant can be written as
\beq
y^2(t)=\frac{C-A}{B}g^2(t)\left[1+H_{0}\eta^{1-\frac{C}{A}}(t)\right]^{-1}.
\label{y_nin_cozumu}
\eeq %
 The solution of the third  RG equation
(\ref{5eg2}) can be defined by another RG invariant $P(t)$ if and
only if the constants $B$ equals to $D$ and $C$ equals to $E$.
Then the invariant becomes
\beq
P(t)=-\eta^{-1+C/A}(t)\left[1-\frac{B}{C-A}\frac{y^2(t)}{g^2(t)}\right]
\frac{e^2(t)}{y^2(t)}\frac{g^2(t)}{y^2(t)}.
\eeq %
The solution of the composite vector coupling $e(t)$ can be
written as
\beq
e^2(t)=-\frac{P_{0}}{H_{0}}\left(\frac{C-A}{B}\right)^{2}g^2(t)\left[1+H_{0}
\eta^{1-\frac{C}{A}}(t)\right]^{-1}. \label{e_nin_cozumu}
\eeq %
where $P_{0}$ denotes the value of the invariant $P(t)$. The
solution of the last RG equation (\ref{5L4}) can be defined by
another RG invariant $K(t)$,  given as
\beq %
  K(t)=-\eta^{-1+\frac{2C}{A}}(t)\left[
  1-\frac{2C-A}{2B}\frac{a(t)}{y^2(t)}\frac{g^2(t)}{y^2(t)}\right].
\eeq %
We can rewrite the solution with a value of the invariant $K(t)$
as $K_{0}$
\beq %
 a(t)=\frac{2(C-A)^{2}}{(2C-A)B}g^{2}(t)
 \frac{1+K_{0}\eta^{1-\frac{2C}{A}}(t)}{\left[1+H_{0}\eta^{1-\frac{C}{A}}(t)\right]^{2}}
\label{a_nin_cozumu}.%
\eeq
Here we notice that the RG constants $H_{0}$, $P_{0}$ and $K_{0}$
play important roles on the behavior of the solutions of the
coupling equations (\ref{g_nin_cozumu}), (\ref{y_nin_cozumu}),
(\ref{e_nin_cozumu}), (\ref{a_nin_cozumu}). Similar works have
been studied in \cite{ha_ki_ku_na_94,ho_lu_07}. The values of the
constants are given in these equations as %
\beq %
    A=\frac{11N_C-4T(R)N_f}{3}, \hspace{5mm} B=D=\frac{G}{4}=2n_fN_C,
       \hspace{5mm} C=E=6C_2(R), \hspace{5mm} F=G. %
\eeq %
Here $C_2(R)$ is a second Casimir,
$C_2(R)=\frac{(N_{C}^{2}-1)}{2N_{C}}$, $R$ is the fundamental
representation with $T(R)=\frac{1}{2}$.

Before entering the analysis of the fixed point, we briefly
investigate the results of some limits.
\subsubsection{The limiting case A$\rightarrow$$+0$ for finite $t$}
In this case the coupling constants solutions can be written as
\beq g^2(t)&=&g^2_0, \\
     y^2(t)&=&\frac{8\pi^{2}}{B}\frac{\alpha}{\alpha_{c}}
          \left[1+H_{0}exp\left(\frac{\alpha}{\alpha_{c}}t\right)\right]^{-1},\\
     e^2(t)&=&-\frac{16\pi^{3}}{B^{2}}\frac{P_{0}}{H_{0}}\frac{\alpha}{\alpha_{c}^{2}}
          \left[1+H_{0}exp\left(\frac{\alpha}{\alpha_{c}}t\right)\right]^{-1},\\
     a(t)&=&\frac{8\pi^2}{B}\frac{\alpha}{\alpha_{c}}
          \frac{\left[1+K_{0}exp\left(\frac{2\alpha}{\alpha_{c}}t\right)\right]}
          {\left[1+H_{0}exp\left(\frac{\alpha}{\alpha_{c}}t\right)\right]^{2}}.
\eeq %
Here $\alpha_{0}=\alpha$ and $\frac{C}{2\pi}=\frac{1}{\alpha_c}$.
\subsubsection{The limiting case A$\rightarrow$$C$ for finite $t$}
In this limit case the solutions of the couplings
(\ref{y_nin_cozumu}), (\ref{e_nin_cozumu}) and
(\ref{a_nin_cozumu}) seem to vanish. If we suggest new RG
invariant $H_{1}$, instead of $H_{0}$, as
$H_{0}=-1+\frac{C-A}{A}H_{1}$ , we find that two of the coupling
solutions do not vanish, whereas composite vector coupling goes to
zero.
These behaviors are given below %
\beq
   y^{2}(t)&=&\frac{A}{B}g^{2}(t)\left[H_{1}+\ln\eta(t)\right]^{-1}, \\
   e^{2}(t)&=&P_{0}\left(\frac{C-A}{B}\right)\left(
              \frac{A}{B}\right)g^{2}(t)\left[H_{1}+\ln\eta(t)\right]^{-1}, \\
   a(t)&=& \frac{2A}{B}g^{2}(t) \frac{1+K_{0}\eta^{-1}(t)}{
          \left[H_{1}+\ln\eta(t)\right]^{2}}.
\eeq
It is amusing to see that the added interactions nullify the
original vector-spinor coupling.
\subsubsection{The limiting case A$\rightarrow$$2C$ for finite $t$}
In this limit case only the quartic coupling constant solution
(\ref{a_nin_cozumu}) behaves critically. Similarly we can redefine
RG invariant $K_{1}$ instead of $K_{0}$ as
$K_{0}=-1+\frac{2C-A}{A}K_{1}$, then the quartic coupling solution
takes the form
\beq %
   a(t)=\frac{C}{B}g^{2}(t)\frac{K_{1}+\ln\eta(t)}{
          \left[1+H_{0}\eta^{1/2}(t)\right]^{2}}.
\eeq %
This limit is not allowed because it does not give asymptotic
freedom.

In the next section we will mention which criteria are needed to
define a nontrivial theory.
\section{Nontriviality of the system} \label{nontriviality}
To have a nontrivial theory all the running coupling constants
should not diverge at any finite energy, which means the absence
of Landau poles of the system. For a consistent theory these
solutions should not vanish identically and must have real and
positive values. These conditions make the model unitary and
satisfy the vacuum stability criterion. Note that if we decouple
the scalar and composite vector field from the system, we have a
nontrivial theory, similar to QCD. Therefore, $e(t)\equiv
g(t)\equiv a(t)\equiv 0$ solution will not be named as the
nontriviality of our composite model. The mass parameter can be
renormalized in the $\overline{MS}$ scheme and the mass can be
chosen as zero.

Remember that we are restricted by neglecting the scalar loop
contributions to the gauge coupling where the composite vector
contributions are not neglected but down due to the presence of
$\epsilon$ in its propagator. If the Yukawa and/or quartic scalar
couplings become so large and break the $1/N_{C}$ expansion then
the behavior of the gauge coupling might be affected.

These restriction conditions are the same as the ones in the gHY
system which was discussed widely in \cite{ha_ki_ku_na_94}. A
while ago, one of us, B.C.L., with a collaborator, studied the
scalar form of the G\"ursey model in this fashion \cite{ho_lu_07}.
In that model, there is a composite scalar field with a propagator
completely different from a constituent scalar field used in
reference \cite{ha_ki_ku_na_94}. There, we showed that a
restriction is not needed between the scalar and the gauge field
coupling since the contribution of the scalar field to the gauge
field is down by the factor of $\epsilon$ in the scalar
propagator. In this work, the vector form of the G\"ursey Model,
we have constituent scalar field and composite vector field which
is missing in gHY system. This composite field adds a new RGE to
the system but does not contribute to the former ones in gHY
system with a totally different reason.

After these remarks  we will discuss the nontriviality conditions
of our model in the following subsections.
\subsection{Fixed Point Solution}
The RGE's can be rewritten as %
\beq %
   8\pi^{2}\frac{d}{dt}\left[\frac{y^{2}(t)}{g^{2}(t)}\right]&=&
      Bg^{2}(t)\left[\frac{y^{2}(t)}{g^{2}(t)}\right]
       \left[\frac{y^{2}(t)}{g^{2}(t)}-\frac{C-A}{B}\right], \\
   8\pi^{2}\frac{d}{dt}\left[\frac{e^{2}(t)}{y^{2}(t)}\frac{g^{2}(t)}
   {y^{2}(t)}\right]&=&(C-A)g^{2}(t)\left[\frac{e^{2}(t)}{y^{2}(t)}\right]
       \left[\frac{g^{2}(t)}{y^{2}(t)}-\frac{B}{C-A}\right], \\
   8\pi^{2}\frac{d}{dt}\left[\frac{a(t)}{y^{2}(t)}\frac{g^{2}(t)}
      {y^{2}(t)}\right]&=&(2C-A)g^{2}(t)
      \left[\frac{a(t)}{y^{2}(t)}\frac{g^{2}(t)}{y^{2}(t)}
      -\frac{2B}{2C-A}\right].
\eeq
The fixed point solutions can be given as %
\beq %
   \frac{y^{2}(t)}{g^{2}(t)}&=&\frac{C-A}{B}, \\
   \frac{e^{2}(t)}{y^{2}(t)}&=&\mbox{Arbitrary constant}, \\
   \frac{a(t)}{y^{2}(t)}&=&\frac{2(C-A)^{2}}{B(2C-A)}.
\eeq%
These are also the solutions of the equations
(\ref{y_nin_cozumu}), (\ref{e_nin_cozumu}) and
(\ref{a_nin_cozumu}) where the RG invariants are
$P_{0}=H_{0}=K_{0}=0$ as $P_{0}=\zeta H_{0}$. Here $\zeta$ is a
constant.
It is clear that the behavior of all the coupling constants are
determined by the gauge coupling which means the Kubo, Sibold and
Zimmermann's "coupling constant reduction" \cite{ku_si_zi_89}.
This corresponds to the Pendleton-Ross  fixed point
\cite{pe_ro_81} in the context of the RGE. Remark that only the
case, $C>A$, prevents the violation of the unitarity and keeps the
stability of the vacuum. This gives rise to nontriviality of the
model when the RG invariants are set to zero. In the following
subsections we will analysis the coupling constant solutions only
in this case with non zero RG invariants.
\subsection{Yukawa Coupling}
The Yukawa coupling solution is given in equation
(\ref{y_nin_cozumu}). It is obvious that the sign of the RG
invariant, $H_{0}$, plays an important role in the behavior of the
solution where B is positive. The ultraviolet (UV) limit of
$\eta(t)$ is needed before continuing the analysis in $C>A$ case.
\beq %
   \eta^{1-\frac{C}{A}}(t \rightarrow\infty)\rightarrow \ %
 \begin{array}{ll}
  +\infty. \\
  \end{array}%
\eeq %
The UV behavior of Yukawa coupling with a non zero RG invariant
$H_{0}$ is
\beq %
   y^2(t\rightarrow\infty)\rightarrow\ \left\{%
\begin{array}{lrl}
    +0, \hspace{5mm} &\hbox{$0<$}&\hbox{$H_{0}<\infty$;} \\
    \hbox{Landau Pole}, &\hspace{5mm} \hbox{$-1<$}&\hbox{$H_{0}<0$;} \\
    -0, \hspace{5mm} &\hbox{$-\infty<$}&\hbox{$H_{0}\leq-1$.} \\
\end{array}%
\right. %
\eeq %
For $-1<H_{0}<0$ case, there exists a finite value of $t$ before it goes to infinity %
\beq %
   1+\frac{A\alpha_{0}}{2\pi}t=\left(\frac{-1}{H_{0}}\right)^{A/(C-A)}.
\eeq %
In this $t$ value Yukawa coupling diverges and changes its sign.
These asymptotic behaviors show that the theory is nontrivial if
and only if the RG invariant $H_{0}$ is positive.

\begin{figure}[htb!]
   \epsfxsize=85mm %
   \epsffile{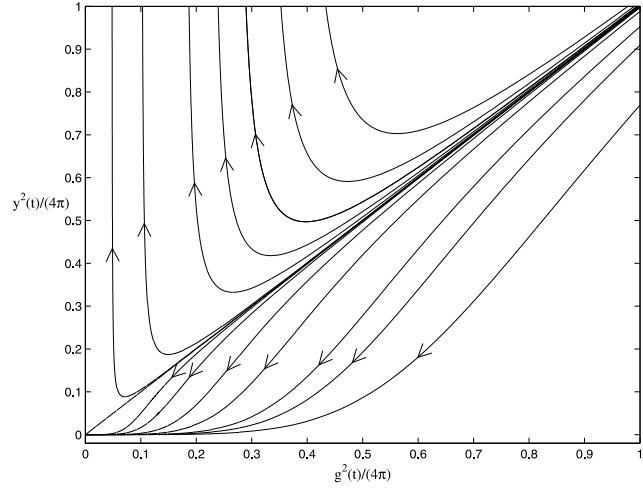} %
   \caption{Plot of $g^2(t)$ vs. $y^2(t)$ for different values of
 $H_0$. The arrows denote the flow directions toward the UV
 region.} \label{y_kare_g_kare}%
\end{figure}

The RG flows in the $(g^{2}(t),y^{2}(t))$ plane are shown in
Figure \ref{y_kare_g_kare}. The upper bound of the figure denotes
the "Landau Pole".
%
%
\subsection{Composite Vector Field Coupling}
\begin{figure}[htb!]
   \epsfxsize=85mm %
   \epsffile{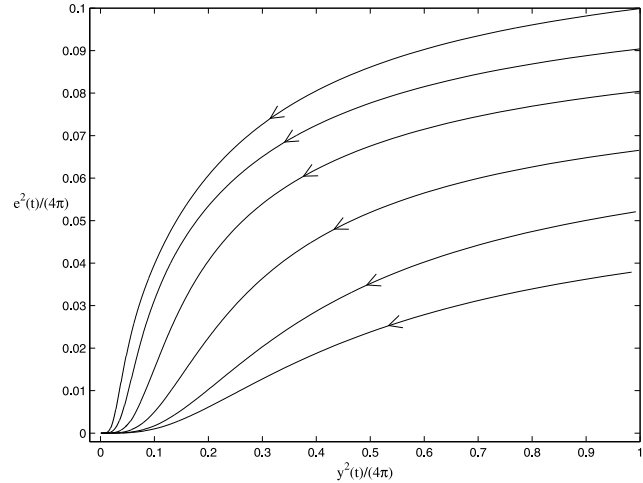} %
   \caption{Plot of $y^2(t)$ vs. $e^2(t)$ for the values of
      $H_{0}>0$ and $P_{0}<0$.}
   \label{e_kare_y_kare}%
\end{figure}

The composite vector coupling solution is given in equation
(\ref{e_nin_cozumu}). In this case not only the sign of $H_{0}$
but also the sign of $P_{0}$ is crucial for nontriviality. Since
$H_{0}$ is positive, $P_{0}$ must be negative. The composite
vector field coupling behaves similarly to the Yukawa coupling up
to a constant multiplier. In figure \ref{e_kare_y_kare} we plot
$e^2(t)$ vs. $y^2(t)$ where $P_{0}<0$, $H_{0}>0$.  Both coupling
constants approach the origin as $t$ goes to infinity. Thus, our
model fulfills the condition required by the asymptotic freedom
criterion.
%
%
\subsection{Quartic Scalar Field Coupling}
\begin{figure}[htb!]
   \epsfxsize=85mm %
   \epsffile{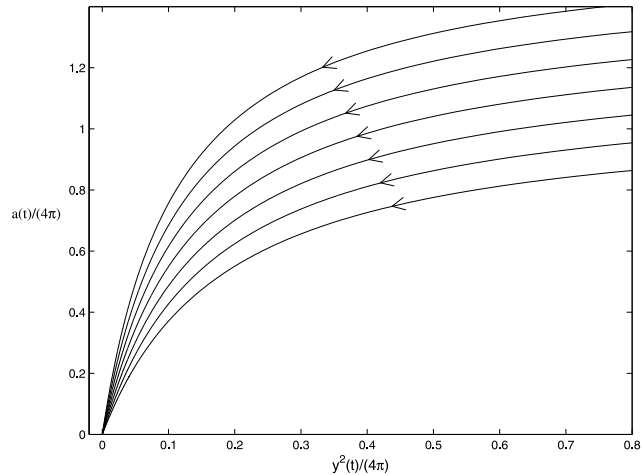} %
   \caption{Plot of $a(t)$ vs. $y^2(t)$ for the values of
      $H_{0}>0$ and $K_{0}=0$.}
   \label{a_y_kare}%
\end{figure}

Finally quartic scalar coupling solution given in equation
(\ref{a_nin_cozumu}) can be analyzed. We have already restricted
ourselves with $C>A$, $H_{0}>0$ and $P_{0}<0$ for nontriviality.
In the limit where $t\gg 1$ , the $\eta$ terms in the last
fraction of equation (\ref{a_nin_cozumu}) become dominant
therefore $1$ can be neglected. Hence we can express the solution
as
\beq
   a(t)\approx \frac{2(C-A)^{2}}{(2C-A)B}g_{0}^{2}\eta(t) \frac{K_{0}\eta^{1-2C/A}(t)}{
          \left[H_{0}\eta^{1-C/A}(t)\right]^{2}},
\eeq %
which is equal to
\beq
   a(t\rightarrow\infty)=\frac{2(C-A)^{2}}{(2C-A)B}g_{0}^{2} \frac{K_{0}}{H_{0}^{2}}.
\eeq %
This asymptotic behavior shows that to have a nontrivial model the
RG invariant $K_{0}$ should be equal to zero. The other
possibilities for a non zero solution for $K_{0}$ is been widely
discussed in the reference \cite{ha_ki_ku_na_94}. In Figure
\ref{a_y_kare}, we plot the RG flows in $(a(t),y^2(t))$ plane for
different values of $H_0$ higher than zero while the gauge
coupling $\alpha(t=0)$ is fixed to one. The origin is the limit
where $t$ goes to infinity, there both
coupling constants approach zero when $K_{0}=0$. %
%
%
\section{Conclusion}
A while ago, one of us, F.T., with a collaborator, showed that the
scattering of composite vector particles gives nontrivial results
while the constituent spinors do not. In that work
\cite{ho_ta_07}, a polynomial Lagrangian model inspired by the
vector form of G\" ursey model was used. Here we couple a
constituent massless scalar field to our previous model. We find
out that many of the features, related to the creating and
scattering of the spinor particles of the original model, are not
true anymore. In the one loop approximation we find the RGE's
whose solutions have all the problems associated with the Landau
pole, like the case in reference \cite{ho_lu_ta_07}. To remedy
this defect we couple a $SU(N_{C})$ non-Abelian gauge field to the
new model. We solve the new RGE's and conclude that if the
conditions $C>A$, $H_{0}>0$, $P_{0}\leq 0$ and $K_{0}=0$ are
satisfied, the model gives a result which can be interpreted as a
nontrivial field theoretical model. We find fixed point solutions
where the coupling constants are not equal to zero. In section
\ref{nontriviality} we plot the UV region behavior of the coupling
constants. There, they all go to zero asymptotically which means
asymptotic freedom, which is another feature of a nontrivial
model.

Our calculation shows that one can construct nontrivial field
theory starting from constrained Lagrangians.

\vspace{5mm}\textbf{Acknowledgement}: We thank to Mahmut Horta\c
csu for discussions and both scientific and technical assistance
while preparing this manuscript. We also thank Nazmi Postac\i
o\={g}lu for technical discussions. This work is supported by the
ITU BAP project no: 31595. This work is also supported by TUBITAK,
the Scientific and Technological Council of Turkey.

\end{document}